 \documentclass[aps,prl,twocolumn,showpacs,10pt,superscriptaddress,floatfix]{revtex4-1}

\usepackage{graphicx}
\usepackage{subfigure}
\usepackage{calrsfs}
\DeclareMathAlphabet{\pazocal}{OMS}{zplm}{m}{n}
\usepackage{amsmath}
\usepackage{amsfonts}
\usepackage{amssymb}
\usepackage{mathrsfs}
\usepackage{bm}
\usepackage{subfigure}
\usepackage{xcolor}
\usepackage[colorlinks,citecolor=blue,linkcolor=red]{hyperref}

\newlength{\seplinewidth}
\newlength{\seplinesep}
\colorlet{sepline}{orange}

\begin{document}

\title{Dark Matter Detection with Strongly Correlated Topological  Materials: Flatband Effect}

\author{Zhao Huang}
\affiliation{Theoretical Division, Los Alamos National Laboratory, Los Alamos, New Mexico 87545, USA}

\author{Christopher Lane}
\affiliation{Theoretical Division, Los Alamos National Laboratory, Los Alamos, New Mexico 87545, USA}
\affiliation{Center for Integrated Nanotechnology, Los Alamos National Laboratory, Los Alamos, New Mexico 87545, USA}

\author{Sarah E. Grefe}
\affiliation{Theoretical Division, Los Alamos National Laboratory, Los Alamos, New Mexico 87545, USA}

\author{Snehasish Nandy}
\affiliation{Theoretical Division, Los Alamos National Laboratory, Los Alamos, New Mexico 87545, USA}
\affiliation{Center for Nonlinear Studies, Los Alamos National Laboratory, Los Alamos, NM, 87545, USA}

\author{Benedikt Fauseweh}
\affiliation{Theoretical Division, Los Alamos National Laboratory, Los Alamos, New Mexico 87545, USA}
\affiliation{German Aerospace Center (DLR), 51147 Cologne, Germany}

\author{Silke Paschen}
\affiliation{Institute of Solid State Physics, Vienna University of Technology, 1040 Vienna, Austria}

\author{Qimiao Si}
\affiliation{Department of Physics \& Astronomy, Center for Quantum Materials, Rice University, Houston, Texas 77005, USA}

\author{Jian-Xin Zhu}
\email{jxzhu@lanl.gov}
\affiliation{Theoretical Division, Los Alamos National Laboratory, Los Alamos, New Mexico 87545, USA}
\affiliation{Center for Integrated Nanotechnology, Los Alamos National Laboratory, Los Alamos, New Mexico 87545, USA}

\begin{abstract}
Dirac materials have been proposed as a new class of electron-based detectors for light dark-matter (DM) scattering or absorption, with predicted sensitivities far exceeding superconductors and superfluid helium. The superiority of Dirac materials originates from a significantly reduced in-medium dielectric response winning over the suppression of DM scattering owing to the limited phase space at the point-like Fermi surface. Here we propose a new route to enhance significantly the DM detection efficiency via strongly correlated topological semimetals. Specifically, by considering a strongly correlated Weyl semimetal model system, we demonstrate that the strong correlation-induced flatband effects can amplify the coupling and detection sensitivity to light DM particles by expanding the scattering phase space, while maintaining a weak dielectric in-medium response. 
\end{abstract}
\date{\today}
\maketitle

{\it Introduction.---} 
Dark matter (DM) makes up the majority of the total matter in the universe, and yet has been difficult to detect. In recent years there is a growing interest in identifying powerful targets for detection. Direct identification of DM particles relies on observing the response of a target to its interaction with the DM through the scattering or absorption process. This in turn depends on the DM energy deposited on the target. Current experimental techniques that can probe the  DM in the mass range of 10 GeV to 10 TeV~\cite{EAprile:2012,DAkerib:2014,RAgnese:2014} are based on collision induced nuclei recoils. Because  the electron mass is much smaller than its nuclear counterpart, electron-based targets enable the detection of much lighter DM particles. Electron materials with an excitation energy of 1 eV order of magnitude are optimal for the detection of DM in the 1 MeV to sub-GeV mass range via direct scattering processes~\cite{REssig:2012a,REssig:2012b,REssig:2016,PWGraham:2012,YHochberg:2017a,SDerenzo:2017,REssig:2017,RBudnik:2018}, or bosonic DM with mass greater than 1 eV via absorption processes~\cite{YHochberg:2017b,IMBloch:2017}. For direct detection of even lighter DM (e.g., particles with masses below MeV through scattering and below 1 eV through absorption), a material with an even narrower  excitation energy gap of few meV or a few decades of meV is desirable. In this regime, superconducting~\cite{YHochberg:2016a,YHochberg:2016b,YHochberg:2016c} and superfluid helium~\cite{KSchutz:2016,SKnapen:2017} have been proposed as possible targets, but each lacks optimal sensitivity due to their larger in-medium response. 

More recently, Dirac materials~\cite{NPArmitage:2018} have attracted enormous attention~\cite{YHochberg:2018,ACoskuner:2019,RMGeihufe:2020} for being a new class of electron-based targets for light DM detection, due to their superior sensitivity compared to superconductors and superfluid helium. Here, a significantly reduced in-medium dielectric response wins over the suppression of DM scattering owing to the limited phase space at the point-like Fermi surface in these materials.
However, while the community has been focusing on weakly interacting Dirac semimetals, a key barrier to progress is to identify new mechanisms that promote enhanced sensitivity to the dark matter. 
In this Letter, we propose to utilize topological quantum materials with strong correlations to fill this void. Our work takes advantage of substantial recent advances in Weyl semimetals driven by strong correlations, especially Weyl-Kondo semimetals~\cite{SDzsaber:2017,HHLai:2018,SDzsaber:2021}.
Starting with a strongly correlated Weyl semimetal model system, we reveal that the correlation effects can significantly magnify the coupling or detection sensitivity to light DM particles. This enhancement is associated with the strong correlation-induced flatband effect due to band renormalization, which not only enhances the scattering phase space but also retains the reduced in-medium effect.


{\it Correlation-Driven Band Renormalization and Topological Phase Transition in Weyl Semimetal Model.---} As a proof-of-principle, we consider a minimal model of a topological Weyl-Hubbard semimetal  defined on a three-dimensional (3D) cubic lattice. 
Within the Gutzwiller projected wavefunction
method~\cite{MCGutzwiller:1963,CLi:2006,JPJulien:2006,QHWang:2006,WHKo:2007,NFukushima:2008,JXZhu:2012}, the Weyl-Hubbard model Hamiltonian reduces to:
\begin{eqnarray}
\label{eq:Hamil2}
H&=&\alpha  \sum_{i,ss^{\prime}} \biggl{[}  -t  \sigma_{x,ss^{\prime}} (c_{js}^{\dagger}   c_{j+\hat{x},s^{\prime}} 
+c_{js}^{\dagger}   c_{j+\hat{y},s^{\prime}} + c_{js}^{\dagger}   c_{j+\hat{z},s^{\prime}} ) \nonumber  \\
&& -i t^{\prime}  ( \sigma_{y,ss^{\prime}}    c_{js}^{\dagger}   c_{j+\hat{y},s^{\prime}}
 +  \sigma_{z,ss^{\prime}}   c_{js}^{\dagger}   c_{j+\hat{z},s^{\prime}}  )  + \text{H.c.} \biggr{]}   \nonumber  \\
&& + m \sum_{j,ss^{\prime}}  \sigma_{x,ss^{\prime}} c_{js}^{\dagger}   c_{js^{\prime}} + Ud N_L\;.
\end{eqnarray}
Here $t$, $t^{\prime}$ are the hopping integrals, $m$ is the strength of an on-site effective in-plane spin Zeeman energy, and $U$ is the Hubbard interaction between two electrons of opposite spin directions on the same site. 
The parameters $\alpha$ and $d$ are the renormalization factor and double occupancy subject to self-consistency conditions, respectively, and $N_L$ is the number of 3D lattice sites, see the Supplemental Material (SM)~\cite{SupplementaryInformation} for more details. In the following discussions, energy is measured in units of $t$, and length in units of cubic lattice constant $a$. Both $t$ and  $a$ are assumed to be one unless specified otherwise. We further fix $t^{\prime}=0.2$, and $m=0.125$.  

\begin{figure}
\centering
\includegraphics[width=1\linewidth]{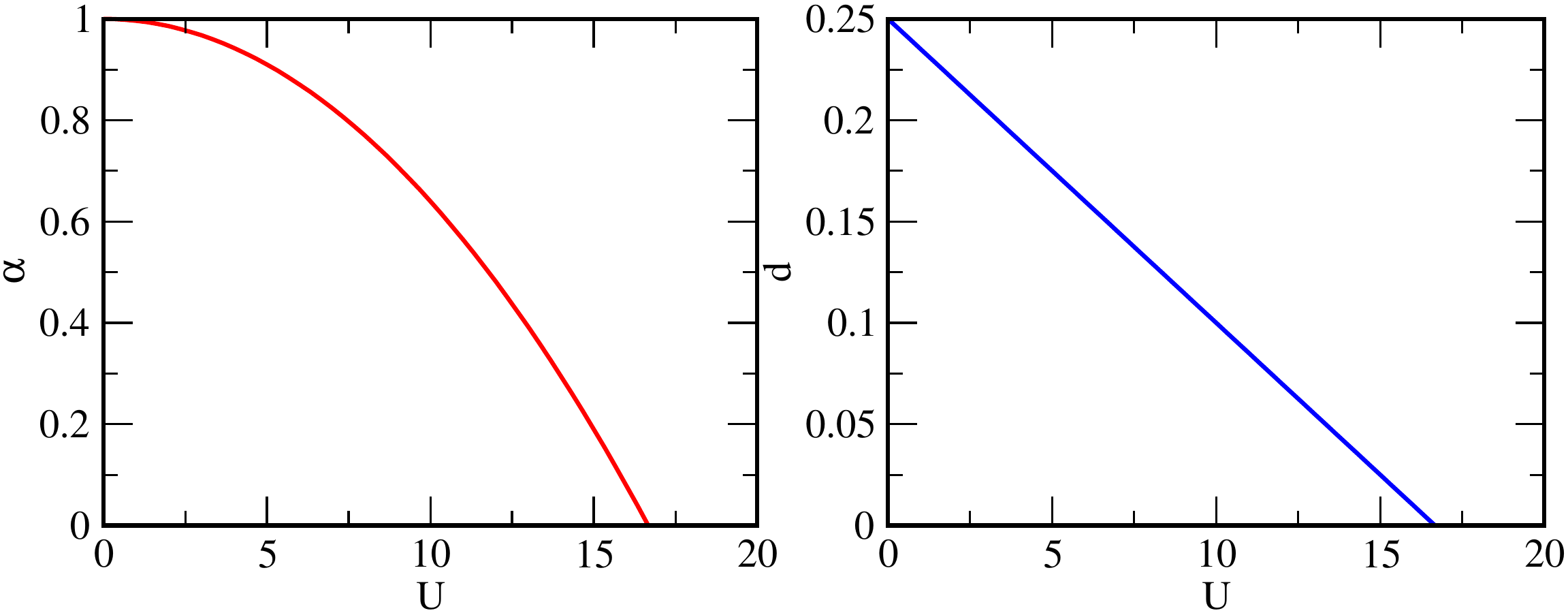}
\caption{Hubbard-$U$ dependence of the hopping renormalization parameter $\alpha$ (a) and double occupancy parameter $d$ (b).}
\label{Fig:Gutz}
\end{figure}

Figure~\ref{Fig:Gutz} shows the variation of $\alpha$ and $d$ as a function of the Hubbard-$U$ strength. 
For $U=0$, the system is non-interacting and yields the well known values $\alpha=1$ and $d=1/4$. Both $\alpha$ and $d$ decrease with increasing value of $U$ and vanish at the Brinkman-Rice (BR)~\cite{WFBrinkman:1970} transition point for $U_c=16.67$. We caution that while the BR picture based on the Gutzwiller approximation gives a physically transparent description of the Mott metal-insulator transition for a half-filled particle-hole symmetric single-band Hubbard model, the BR transition itself is an artifact of the infinite-dimension limit~\cite{PvanDongen:1989,FGebhard:1990}. 
However, since the purpose of the present work is focused on capturing correlation-driven band structure renormalization effects and associated topological phase transitions, which occur before the BR transition, 
 we anticipate the consequence of flat electron bands on dark matter detection to be robust especially for the lighter dark matter, for which the quasiparticle excitions are long-lived.

\begin{figure}
\centering
\includegraphics[width=1\linewidth]{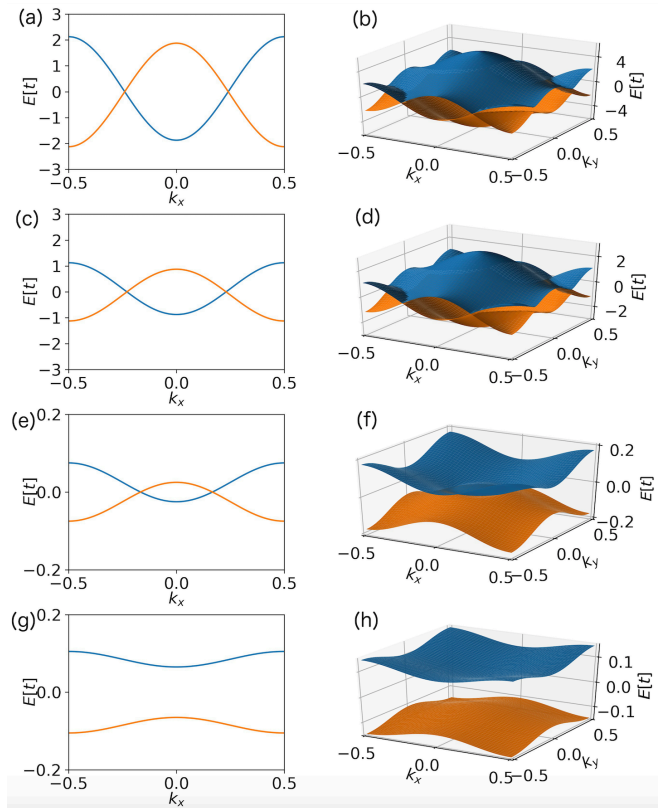}
\caption{2D and 3D plots for band dispersion for various values of $\alpha=1$ [$U=0$] (a-b), 0.5 [$U=11.7869$] (c-d), 0.025 [$U=16.4595$] (e-f), and 0.01 [$U=16.5857$] (g-h). For the 2D plots, the wavevector path is taken to be parallel to $k_x$-axis at $k_y=\pi$ for (a) and (c), and $k_y=0$ for (e) and (g). 
$k_z$ is fixed to be 0.
}
\label{Fig:ESK}
\end{figure}

\begin{figure*}
\centering
\includegraphics[width=1\linewidth]{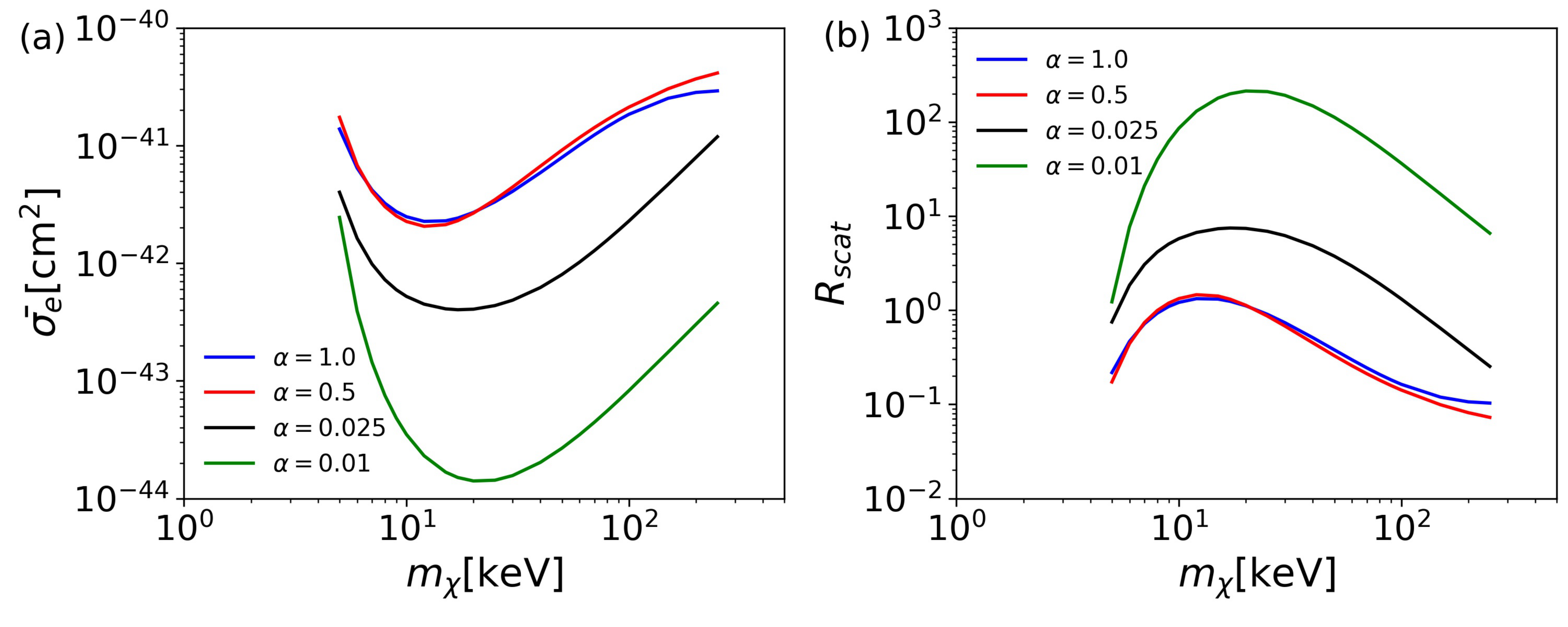}
\caption{Correlation-driven flatband effect on dark matter scattering. (a) Projected reach of dark matter scattering in Weyl semimetals, for varying band renormalization parameter $\alpha=1.0$, 0.5, 0.025, and 0.01 
through a light kinetically mixed dark photon mediator with in-medium effects included. By 
following Ref.~\onlinecite{YHochberg:2018}, we show the expected back-ground-free 95\% confidence level sensitivity (3 events) that can be obtained with 1 kg-yr exposure. (b) Scattering rate as a function of light dark matter mass for varying $\alpha$ at a fiducial cross section of $10^{-42}$.}
\label{Fig:Scattering}
\end{figure*} 

By obtaining the self-consistent Gutzwiller variational parameters, the quasiparticle band dispersion is expressed as 
\begin{equation}
E_{\pm,\mathbf{k}} =\pm \sqrt{\xi_{\mathbf{k}}^2 + \vert \Delta_{\mathbf{k}}\vert^{2}} = \pm E_{\mathbf{k}}^{(0)} \;,
\label{eq:eigenvalues}
\end{equation}
where
$\xi_{\mathbf{k}}    = 2 \alpha t^{\prime} \sin k_z$,  $\Delta_{\mathbf{k}} = Z_{\mathbf{k}} -2i\alpha t^{\prime} \sin k_y$ with 
$Z_{\mathbf{k}} = m-2\alpha t (\cos k_x + \cos k_y + \cos k_z) $.
Figure~\ref{Fig:ESK} displays the electronic band structure along paths parallel to the $k_x$-axis ($[k_x, \pi,0]$ or $[k_x, 0,0]$) in the $k_z=0$ plane for decreasing values of $\alpha$. Since the model described in Eq.~(\ref{eq:Hamil2}) breaks both time-reversal symmetry (through the effective Zeeman term) and inversion symmetry (by the additional $\frac{\pi}{2}$-shifted hopping along the $y$- and $z$-directions), it produces a Weyl semimetallic phase that hosts Weyl nodes with locally linear band dispersions.
In particular, as $U$ increases ($\alpha$ decreases) in strength,  we find the system to undergo two topological phase transitions: WSM phase-I ($\alpha > m/2t$) time-reversal symmetry weakly broken and four Weyl nodes appear at momenta points $(\pm \cos^{-1}(m/2\alpha t), \pi, 0)$ and $(\pm \cos^{-1}(m/2\alpha t), 0, \pi)$. In WSM phase-II ($m/6t <\alpha < m/2t$) time-reversal symmetry is strongly broken, with only two Weyl nodes located at $(\pm \cos^{-1}(m/2\alpha t -2), 0, 0)$ in the Brillouin zone. For $\alpha < m/6t$, all Weyl nodes are gapped out leaving the system in a topologically trivial insulating narrow band phase. Interestingly, the latter transition has recently been observed experiments~\cite{SDzsaber:2022} and appears in a Kondo lattice model~\cite{SGrefe:2020}. The Weyl nature of the semimetallic phases and the topologically trivial insulating phase were determined by analyzing the Berry curvature and $\mathbb{Z}_2$ topological index, as detailed in SM~\cite{SupplementaryInformation}. Note that the electron correlations not only drive topological phase transitions, but also flatten the band dispersion, as also found in experiments~\cite{SDzsaber:2017}.

{\it  Correlation-Driven Flatband Effects on Dark Matter Detection Rates.---} To study the flatband effect on the dark matter detection, we consider both the scattering and absorption mechanisms of dark matter~\cite{REssig:2016,YHochberg:2018,TTrickle:2020,AKDrukier:1986}. The detailed formalism is given in the SM~\cite{SupplementaryInformation}. The central quantity entering into both the scattering and absorption rates of DM is the dynamical momentum dependent dielectric function $\overleftrightarrow{
\bm{\epsilon}}(\mathbf{q},\omega)$.  In existing literature~~\cite{YHochberg:2018,ACoskuner:2019,RMGeihufe:2020}, this key quantity is calculated through the density-density correlator, which is valid for isotropic and weakly anisotropic media. In addition, for the calculation of scattering rates in non-interacting Dirac materials like ZrTe$_5$~\cite{YHochberg:2018}, the material parameters were derived from density functional theory. This procedure is valid only near the Dirac nodes, which limits a realistic treatment of the dielectric function in real materials. In the present work, we propose to evaluate this quantity via the conductivity tensor according to the relation $\epsilon_{\alpha\beta}(\mathbf{q},\omega) = \delta_{\alpha\beta} + 4\pi i \sigma_{\alpha\beta}/\omega$, where $\omega$ is the circular frequency and $\sigma_{\alpha\beta}$ is the matrix element of the conductivity tensor. For the Hamiltonian given in Eq.~(\ref{eq:Hamil2}), the dynamical conductivity tensor is calculated using the current-current correlator through the Kubo formula~\cite{JXZhu:2016} and evaluated by integrating over the entire Brillouin zone; see SM~\cite{SupplementaryInformation} for details. This general formalism not only enables a full description of anisotropic effects in real materials but also allows for the inclusion of correlation-driven band renormalization effects in a transparent way. To discuss the correlation-driven flatband effect from a realistic materials perspective, we choose a nearest-neighbor hopping integral of $t=0.1$ eV and $a=4$ \AA. 

Figure~\ref{Fig:Scattering}(a) shows the sensitivity reach projection for DM scattering in a correlated Weyl semimetal through a light kinetically mixed dark photon for various values of the Hubbard interaction. As is plainly seen, the behavior of the fiducial cross section as a function of DM energy sits between a fully gapped standard semiconductor and a perfect Dirac system, since our Weyl semimetal system contains both features in the entire Brillouin zone. Furthermore, our results show that the optimally detectable DM, corresponding to the region of the minimal threshold of the fiducial cross section, is in the range of 10 to 100 keV. The correlation effects do not change notably the detection depth of the DM energy. Instead, they significantly reduces the threshold of the fiducial cross section from $10^{-41}$ cm for smaller $U$-values down to $10^{-44}$ cm for large $U$-values. To further quantify this sensitivity enhancement by correlation effects, Fig.~\ref{Fig:Scattering}(b) shows the DM scattering rate for various Hubbard-$U$ values at a fixed fiducial cross section of $10^{-42}$. It demonstrates that the optimal scattering rate increases from an order of 1 for small $U$-values up to $10^2$ for large $U$-values. This result suggests that correlation-driven flatband Weyl semimetals or semiconductors have significant advantages not only for maximal phase space availability as in conventional metals and the reduced optical response as in semiconductors, but also for very narrow ``effective'' band gap, which can be used to optimize the DM scattering sensitivity by reducing the in-medium effect. 

\begin{figure*}
\centering
\includegraphics[width=1\linewidth]{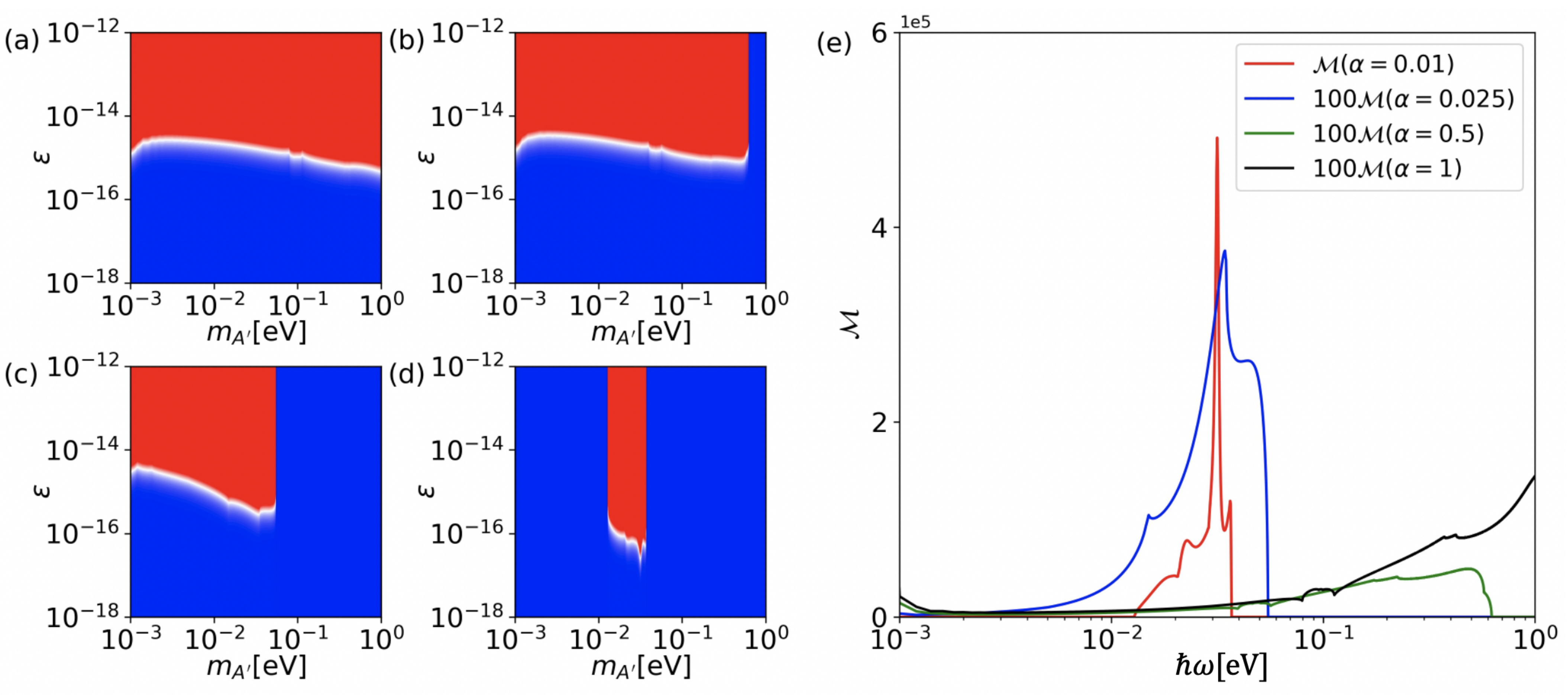}
\caption{Correlation-driven flatband effect on dark matter absorption. (a-d) Projected reach of kinetically mixed dark photon absorption in Weyl semimetals, for varying band renormalization parameter $\alpha=1.0$, 0.5, 0.025, and 0.01. 
It is given in terms of the parameter of $\varepsilon$ for the kinetic mixing between photon and dark photon~\cite{HAn:2013}. By following 
Ref.~\onlinecite{YHochberg:2018}, we show the expected back-ground-free 95\% confidence level sensitivity (3 events) that can be obtained with 1 kg-yr exposure. (b) Dark photon independent target specific absorption efficiency as a function of dark photon energy for varying band renormalization parameter $\alpha$.}
\label{Fig:Absorption}
\end{figure*}

For the detection of kinetically mixed dark photons via the absorption mechanism, where the momentum transfer due to the deposited DM particles is much smaller than their energy (or equivalently their mass), a vertical transition between the valence band and conduction band can occur. However, because the effective in-medium mixing angle between dark and regular photons involves both real and imaginary parts of the polarization tensor (related to the dielectric tensor), the absorption probability is proportional to the ratio, $\frac{\text{Im}\epsilon(m_A^{\prime})}{\vert \epsilon(m_A^{\prime}) \vert^2}$, where $m_A^{\prime}$ is the mass of the kinetically mixed dark photon. This indicates the DM absorption process is more complicated than the regular optical absorption process in a material.   

Figure~\ref{Fig:Absorption}(a-d) shows the correlation effects on the projected sensitivity reach of the Weyl semimetal for the direct absorption of kinetically mixed dark photons. By comparing the projected depth for different values of the band renormalization $\alpha$ (controlled by the Hubbard-$U$ interaction), it is clear that correlation effects do not impact significantly the depth of the kinetic mixing parameter $\varepsilon$ in contrast to DM scattering processes. Instead they reduce the upper bound of the detectable DM photons from the order of 1 eV for weak interactions down to 100 meV for strong interaction strengths. Additionally, the lower bound of DM photon masses is pushed down to 1 meV in the Weyl semimetal phases due to the presence of Weyl nodes irrespective of interaction strength. In the insulating phase ($\alpha=0.01$), the appearance of a narrow semiconducting gap produces a finite lower bound threshold of DM photons around 10 meV. To elucidate this behavior, we evaluate the absorption efficiency as defined by the ratio $\mathcal{M}=\sum_{n} \frac{\text{Im}\epsilon_{n}(m_A^{\prime})}{\vert \epsilon_{n}(m_A^{\prime}) \vert^2}$ with $n$ being the eigen index of the dielectric tensor for the Weyl-Hubbard system. Here, we find the maximal efficiency to be governed by a delicate balance between the imaginary part and the norm squared of the dielectric function. Figure~\ref{Fig:Absorption}(e) shows the energy dependence of $\mathcal{M}$ for various values of renormalization parameter $\alpha$. As shown, the efficiency $\mathcal{M}$ is bounded by the effective band width of the material. In the Weyl semimetal phases, it exhibits a tail behavior as $\omega \rightarrow 0$; while in the insulating phase, it exhibits a gap-like nature in the low frequency region. Noticeably, due to the flattened band dispersion, the efficiency intensity is dramatically increased in WSM phase-II and the insulating phase, which significantly enhances the sensitivity.   

{\it Concluding remarks.---} In conclusion, we have used a model Weyl-Hubbard system as an example for strongly correlated topological quantum matter to investigate the flatband effects on DM detection. 
We have found that with an order of 1 eV of non-interacting electron bandwidth, the strong correlation effects can push deeper the threshold of the fiducial cross section for the optimally detectable DM in the range of 10 to 100 keV through the scattering process while they can tune the detection regime of the dark photons (1 meV to 100 meV) through the absorption process. More importantly, we have discovered that the correlation-driven flatband behavior can significantly enhance the DM detection sensitivity in both scattering and absorption mechanisms. 
In addition, our results suggest, for direct absorption of kinetically mixed dark photons, one can use flatband features as a design principle for constructing highly sensitive DM detectors with selective dark photon energy regimes. In real materials, it is required that the low-energy correlated bands are well separated from high-energy bands. Recently, the nontrivial band topology and strong correlation-driven flatbands were observed in $\rm Ce_3Pd_3Bi_4$~\cite{HHLai:2018,CCao:2020,SDzsaber:2021,SDzsaber:2022}, which could be a promising candidate material platform for direct experimental verification of our dark matter detection predictions. We comment that although we have focused on the flatband effect driven by strong correlation in a Weyl system on the DM detection, the flatband behavior can also be realized in other settings, for example, quantum Hall systems~\cite{DCTsui:1982}, Kagome~\cite{JWRhim:2021} and twisted bilayer graphene~\cite{YCao:2018} systems and we believe that the findings in the present work might be general. This expanded class of materials may provide a new paradigm for the direct detection of fundamental particles.

{\it Acknowledgments.---} We thank Yonit Hochberg, Felix Kahlhoefer, Yonatan Kahn, Filip Ronning, and Kathryn Zurek for stimulating discussions.  
This   work   was   carried   out   under   the   auspices of  the  U.S.  Department  of  Energy  (DOE)  National Nuclear  Security  Administration  under  Contract  No. 89233218CNA000001. It was supported by the LANL LDRD Program (C.L., S.N., B.F., \& J.-X.Z.), the Center for the Advancement of Topological Semimetals, a DOE BES EFRC (Z.W. \& S.G.). 
S.P. acknowledges funding by the European Union (ERC, CorMeTop, project 101055088). Q.S. acknowledges the primary support of the U.S. DOE BES under Award No.
DE-SC0018197, and by the Robert A. Welch Foundation Grant No. C-1411. S.P and Q.S. acknowledge the hospitality of the KITP, UCSB, where support was provided 
by the National Science Foundation under Grant No. NSF PHY-1748958.  It was also supported in part by the Center for Integrated Nanotechnologies, a DOE BES user facility, in partnership  with  the  LANL  Institutional  Computing  Program for computational resources.


\end{document}


\title{Supplemental Material: Dark Matter Detection with Strongly Correlated Topological  Materials: Flatband Effect}

\author{Zhao Huang}
\affiliation{Theoretical Division, Los Alamos National Laboratory, Los Alamos, New Mexico 87545, USA}

\author{Christopher Lane}
\affiliation{Theoretical Division, Los Alamos National Laboratory, Los Alamos, New Mexico 87545, USA}
\affiliation{Center for Integrated Nanotechnology, Los Alamos National Laboratory, Los Alamos, New Mexico 87545, USA}

\author{Sarah E. Grefe}
\affiliation{Theoretical Division, Los Alamos National Laboratory, Los Alamos, New Mexico 87545, USA}

\author{Snehasish Nandy}
\affiliation{Theoretical Division, Los Alamos National Laboratory, Los Alamos, New Mexico 87545, USA}
\affiliation{Center for Nonlinear Studies, Los Alamos National Laboratory, Los Alamos, NM, 87545, USA}

\author{Benedikt Fauseweh}
\affiliation{Theoretical Division, Los Alamos National Laboratory, Los Alamos, New Mexico 87545, USA}
\affiliation{German Aerospace Center (DLR), 51147 Cologne, Germany}

\author{Silke Paschen}
\affiliation{Institute of Solid State Physics, Vienna University of Technology, 1040 Vienna, Austria}

\author{Qimiao Si}
\affiliation{Department of Physics \& Astronomy, Center for Quantum Materials, Rice University, Houston, Texas 77005, USA}

\author{Jian-Xin Zhu}
\email{jxzhu@lanl.gov}
\affiliation{Theoretical Division, Los Alamos National Laboratory, Los Alamos, New Mexico 87545, USA}
\affiliation{Center for Integrated Nanotechnology, Los Alamos National Laboratory, Los Alamos, New Mexico 87545, USA}

\date{\today}
\maketitle

\newcommand{\beginsupplement}{%
        \setcounter{table}{0}
        \renewcommand{\thetable}{S\arabic{table}}%
        \setcounter{figure}{0}
        \renewcommand{\thefigure}{S\arabic{figure}}%
        \setcounter{section}{0}
        \renewcommand{\thesection}{\Roman{section}}%
        \setcounter{equation}{0}
        \renewcommand{\theequation}{S\arabic{equation}}%
        }
\onecolumngrid
\beginsupplement
\setcounter{secnumdepth}{2}

\section{Gutzwiller approximation to the Weyl-Hubbard Model}

Our starting point is a Weyl-Hubbard model with the corresponding Hamiltonian written as:
\begin{eqnarray}
\label{eq:Hamil}
H&=& \sum_{i,ss^{\prime}} \biggl{[}  -t  \sigma_{x,ss^{\prime}} (c_{js}^{\dagger}   c_{j+\hat{x},s^{\prime}} 
+c_{js}^{\dagger}   c_{j+\hat{y},s^{\prime}} + c_{js}^{\dagger}   c_{j+\hat{z},s^{\prime}} ) 
 -i t^{\prime}  ( \sigma_{y,ss^{\prime}}    c_{js}^{\dagger}   c_{j+\hat{y},s^{\prime}}
 +  \sigma_{z,ss^{\prime}}   c_{js}^{\dagger}   c_{j+\hat{z},s^{\prime}}  )  + \text{H.c.} \biggr{]}   \nonumber  \\
&& + m \sum_{j,ss^{\prime}}  \sigma_{x,ss^{\prime}} c_{js}^{\dagger}   c_{js^{\prime}} + U\sum_{i} n_{i\uparrow} n_{i\downarrow}  \;,
\end{eqnarray}
where $t$, $t^{\prime}$ are the hopping integrals, $m$ is the strength of an on-site effective in-plane spin Zeeman exchange interaction energy, and $U$ is the Hubbard interaction between two electrons of opposite spin directions on the same site. 
The strong correlation effect can be treated by reducing the statistical weight of double occupation in the Gutzwiller projected wavefunction approach~\cite{MCGutzwiller:1963}. In a spatially unrestricted Gutzwiller approximation~\cite{CLi:2006,JPJulien:2006,QHWang:2006,WHKo:2007,NFukushima:2008,JXZhu:2012}, we arrive at the following renormalized mean-field Hamiltonian:
\begin{eqnarray}
\label{eq:Hamil-S}
H &=& \sum_{i,ss^{\prime}} \biggl{[}  -t  \sigma_{x,ss^{\prime}} \sqrt{\alpha_{js}} (\sqrt{\alpha_{j+\hat{x},s^{\prime}}} c_{js}^{\dagger}   c_{j+\hat{x},s^{\prime}} 
+ \sqrt{\alpha_{j+\hat{y},s^{\prime}}} c_{js}^{\dagger}   c_{j+\hat{y},s^{\prime}} + \sqrt{\alpha_{j+\hat{z},s^{\prime}}} c_{js}^{\dagger}   c_{j+\hat{z},s^{\prime}} ) \nonumber \\
&& -i t^{\prime} \sqrt{\alpha_{js}} ( \sigma_{y,ss^{\prime}}  \sqrt{\alpha_{j+\hat{y},s^{\prime}}}  c_{js}^{\dagger}   c_{j+\hat{y},s^{\prime}}
 +  \sigma_{z,ss^{\prime}} \sqrt{\alpha_{j+\hat{z},s^{\prime}}}  c_{js}^{\dagger}   c_{j+\hat{z},s^{\prime}}  )  + \text{H.c.} \biggr{]}  \nonumber \\
&& + m \sum_{j,ss^{\prime}}  \sigma_{x,ss^{\prime}} c_{js}^{\dagger}   c_{js^{\prime}} + U\sum_{i} d_{i} \;,
\end{eqnarray}
where $d_i$ is the double occupation at site $i$, and the renormalization parameter $\alpha_{js}$ is given by  
\begin{equation} 
\sqrt{\alpha_{js}}=\biggl{[} \frac{(\bar{n}_{js}-d_{j})(1-\bar{n}_j-d_j)}{\bar{n}_{js} (1-\bar{n}_{js})} \biggr{]}^{1/2}+ \biggl{[}\frac{d_{j}(\bar{n}_{j\bar{s}}-d_j)}{\bar{n}_{js} (1-\bar{n}_{js})} \biggr{]}^{1/2} \;,
\end{equation}
with $\bar{n}_{js}$ the expectation value of the number operator $n_{js}=c^{\dagger}_{js}c_{js}$ and $\bar{n}_j=\sum_{s} \bar{n}_{js}$.
In a uniform paramagnetic phase, $\bar{n}_{js}=\bar{n}_{s}=\bar{n}_{\bar{s}}$, $d_j=d$ and $\alpha_{js}=\alpha$, we obtain 
 Eq.~(1) in the main text. 
Furthermore, at half filling, $\bar{n}_s=1/2$ and $\bar{n}=1$, which yields  
\begin{equation}
\alpha= 16\biggl{(}\frac{1}{2}-d \biggr{)}d \;.
\end{equation}
The minimization of the total energy of the mean-field Hamiltonian at half filling gives the constraint 
\begin{equation}
\frac{\partial{\alpha}}{\partial d} \frac{\langle H_{\text{kin}} 
\rangle}{N_L} + U =0 \;,
\end{equation}
where the kinetic energy part of the Hamiltonian 
\begin{equation}
H_\text{kin} =  \sum_{i,ss^{\prime}} \biggl{[}  -t  \sigma_{x,ss^{\prime}} (c_{js}^{\dagger}   c_{j+\hat{x},s^{\prime}} 
+c_{js}^{\dagger}   c_{j+\hat{y},s^{\prime}} + c_{js}^{\dagger}   c_{j+\hat{z},s^{\prime}} ) 
 -i t^{\prime}  ( \sigma_{y,ss^{\prime}}    c_{js}^{\dagger}   c_{j+\hat{y},s^{\prime}}
 +  \sigma_{z,ss^{\prime}}   c_{js}^{\dagger}   c_{j+\hat{z},s^{\prime}}  )  + \text{H.c.} \biggr{]}  \;.
\end{equation}
Therefore, the parameters $\alpha$ and $d$ are determined self-consistently by solving the eigen problem with Eq.~(2) in the main text.

\section{Topological analysis of the Weyl-Hubbard model}
Once the self-consistent solution is obtained, we can use the momentum dependent eigenvalues and eigenfunctions to perform the topological analysis. 
The eigenvalues are
\begin{equation}
\label{eq:eigenval-S}
E_{\pm,\mathbf{k}} =\pm \sqrt{\xi_{\mathbf{k}}^2 + \vert \Delta_{\mathbf{k}}\vert^{2}} = \pm E_{\mathbf{k}}^{(0)} \;,
\end{equation}
and the corresponding eigenfunctions are
\begin{equation} 
\label{eq:eigenfuncp-S}
\left( \begin{array}{c}
u_{\mathbf{k},+} \\
v_{\mathbf{k},+}  \end{array} \right)  = \left( \begin{array}{c}   
u_{\mathbf{k}}^{(0)} e^{i\varphi_{\mathbf{k}}}\\
v_{\mathbf{k}}^{(0)} \end{array} \right) \;,
\end{equation} 
for $E_{+,\mathbf{k}}$, while 
\begin{equation} 
\label{eq:eigenfuncm-S}
\left( \begin{array}{c}
u_{\mathbf{k},-} \\
v_{\mathbf{k},-}  \end{array} \right)  = \left( 
\begin{array}{c}   
-v_{\mathbf{k}}^{(0)} e^{i\varphi_{\mathbf{k}}}\\
u_{\mathbf{k}}^{(0)} \end{array}  \right)\;,
\end{equation} 
for $E_{-,\mathbf{k}}$. Here 
$\xi_{\mathbf{k}}    = 2 \alpha t^{\prime} \sin k_z$,  $\Delta_{\mathbf{k}} = Z_{\mathbf{k}} -2i\alpha t^{\prime} \sin k_y$ with 
$Z_{\mathbf{k}} = m-2\alpha t (\cos k_x + \cos k_y + \cos k_z) $ and the phase angle $\tan \varphi_{\mathbf{k}} = - 2\alpha t^{\prime} \sin k_y/ Z_{\mathbf{k}}$, and
\begin{equation}
 \left( \begin{array}{c}   
u_{\mathbf{k}}^{(0)} \\
v_{\mathbf{k}}^{(0)} \end{array} \right) = 
\left( \begin{array}{c} 
\sqrt{\frac{1}{2} \biggl{(} 1 + \frac{\xi_{\mathbf{k}}}{E_{\mathbf{k}}^{(0)}}  \biggr{)}  } \\
\sqrt{\frac{1}{2}\biggl{(} 1 - \frac{\xi_{\mathbf{k}}}{E_{\mathbf{k}}}  \biggr{)} }
\end{array} \right)  \;.
\end{equation} 

\begin{figure}
\centering
\includegraphics[width=1\linewidth]{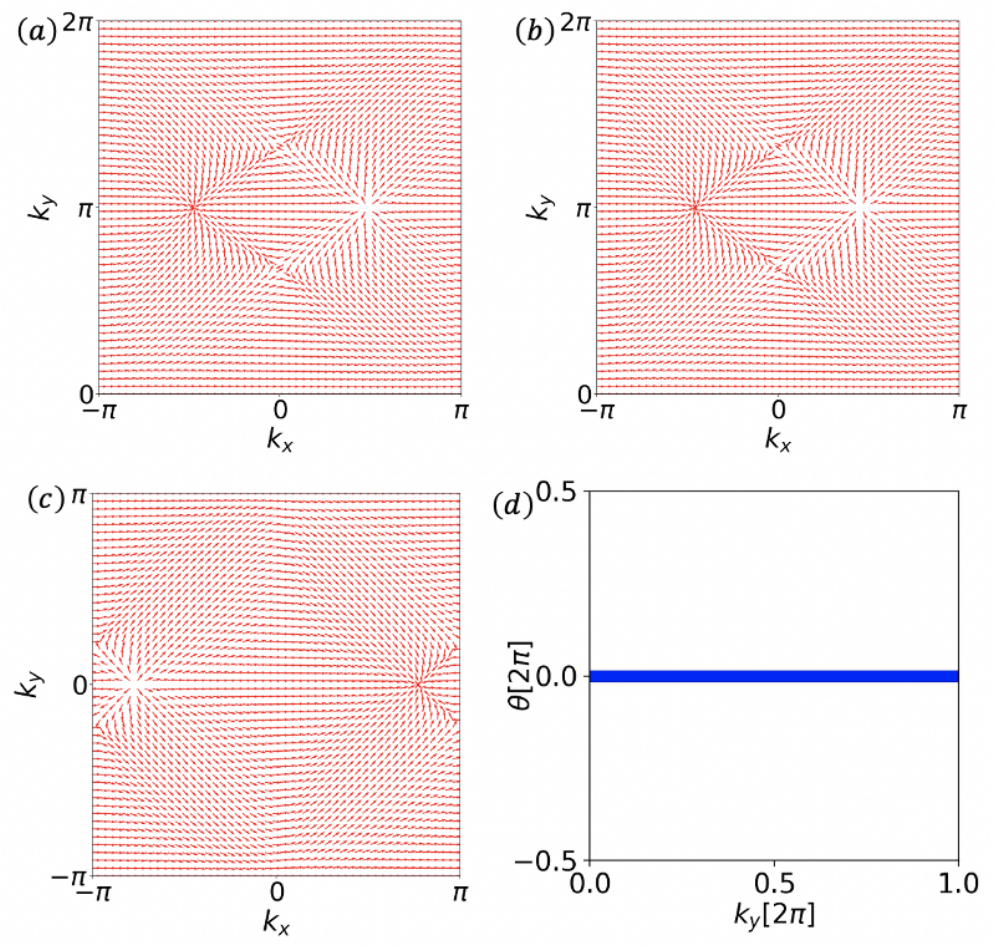}
\caption{Flow of the Berry curvature in the Weyl semimetal phases for $\alpha=1$ (a), 0.5 (b), 0.025 (c), and the evolution lines of the Wannier centers in the insulating phase for $\alpha=0.01$ (d). We choose $k_z=0$.
}
\label{Fig:ESK}
\end{figure}

The Berry curvature vector for band $n$ is given by 
\begin{align}
\Omega_{n,\gamma}(\mathbf{k}) = \varepsilon_{\alpha\beta\gamma} \Omega_{n,\alpha\beta}(\mathbf{k})
\end{align}
with the rank 2 tensor defined as
\begin{align}
\Omega_{n,\alpha\beta}(\mathbf{k})=
-2  \text{Im} \sum_{m\neq n}
\frac{    v_{nm,\alpha}(\mathbf{k})    v_{mn,\beta}(\mathbf{k})   }{    \left[ \varepsilon_{m\mathbf{k}}-\varepsilon_{n\mathbf{k}} \right]^2 }.
\end{align}
where $\varepsilon_{n\mathbf{k}}$ is the eigenvalue of Hamiltonian $H$ for band $n$ and momenta $\mathbf{k}$ and the matrix elements of the Cartesian velocity operators $\hat{v}_{\alpha}=(i/\hbar)\left[ \hat{H},\hat{r}_\alpha \right]$~\cite{DXiao:2010}. The velocity operators can then be straightforwardly obtained, 
\begin{align}
 v_{nm,\alpha}(\mathbf{k})
 &=
\braket{  \phi_{n\mathbf{k}} \left|  \frac{\partial \hat{H}(\mathbf{k})}{\partial k_\alpha}  \right| \phi_{m\mathbf{k}} }\nonumber\\
 &=
\frac{\partial \vec{h}(\mathbf{k})}{\partial k_\alpha}\cdot \braket{  \phi_{n\mathbf{k}} \left| \vec{\mathbf{\sigma}}    \right| \phi_{m\mathbf{k}} }
\end{align}
where $\hat{H}$ is written in vector form,
\begin{align}
    \hat{H}(\mathbf{k})=\begin{bmatrix}
    h_3 & h_1 -i h_2\\
    h_1 +i h_2 &  - h_3 
    \end{bmatrix}
    =\vec{h}(\mathbf{k}) \cdot \vec{\mathbf{\sigma}} 
\end{align}
and
\begin{subequations}
\begin{align}
h_1 &= -2\alpha t_{x} \cos(k_x)-2\alpha t_{y}\cos(k_y)-2\alpha t_{z} \cos(k_z) + m, \\
h_2 &=  2\alpha t^{\prime}_{y}\sin(k_{y}), \\
h_3 &=  2\alpha t^{\prime}_{z}\sin(k_{z}). 
\end{align}
\end{subequations}
Finally, the derivatives with respect to momentum along the various Cartesian directions is then given by,
\begin{subequations}
\begin{align}
\frac{\partial \vec{h}(\mathbf{k})}{\partial k_x} &=  2\alpha t_{x} \sin(k_x)~\hat{x},  \\
\frac{\partial \vec{h}(\mathbf{k})}{\partial k_y} &= 2 \alpha t_{y} \sin(k_y)~\hat{x}+2\alpha t^{\prime}_{y}\cos(k_y) ~\hat{y},\\ 
\frac{\partial \vec{h}(\mathbf{k})}{\partial k_z} &= 2\alpha t_{z}\sin(k_z)~\hat{x}+2\alpha t^{\prime}_{z}\cos(k_z) ~\hat{z}.
\end{align}
\end{subequations}

Figure~\ref{Fig:ESK} (a) - (c) Shows the Berry curvature vector field in the Weyl semimental for $\alpha=1.0$, $0.5$, and $0.25$, respectively. Each panel clearly shows sources (sinks) of Berry curvature indicative of the underlying topologically nontrivial electronic ground state.  Figure~\ref{Fig:ESK} (d) presents the results of the Wilson loop analysis~\cite{RYu:2011} of the trivial $\mathbb{Z}_2$ electronic state for $\alpha=0.01$.

\section{Formalism of dark matter scattering and absorption rates}
For the calculations of dark matter detection rates, we follow closely those in literature~\cite{REssig:2016,YHochberg:2018,TTrickle:2020,AKDrukier:1986}.
The scattering rate is given by
\begin{equation}
R_{scat} = \frac{\rho_{\chi}}{m_{\chi}} \frac{\hbar^2 \pi^2\bar{\sigma}_{e}}{\mu_{\chi e}^2 \mathcal{V}^2\rho_T} \sum_{\mathbf{k}\mathbf{k}^{\prime}\mathbf{G}} \frac{\eta(v_{\rm{min}})\vert \mathbf{q}\vert^3}{ \vert \mathbf{q} \cdot \bm{\epsilon}(\mathbf{q},\omega_{\mathbf{k}\mathbf{k}^{\prime}}) \cdot \mathbf{q}\vert^2}  \vert F_{\rm{DM}}(q)\vert^2  \vert f_{\mathbf{k}, \mathbf{k}^{\prime}}^{-+} \vert^2 \;. 
\end{equation}
Here $\rho_{\chi}$ is the local DM density,  $m_{\chi}$ is the DM mass, $\bar{\sigma}_{e}$ is the fiducial cross section, $\mu_{\chi e}$ is the DM-electron reduced mass, $\rho_T$ is the mass density of the target,  $\mathcal{V}$ is the crystal volume (equal to $N_L a^3$ with $N_L$ being the total number of unit cells in our cubic Weyl-Hubbard model system),  and the momentum transfer $\mathbf{q}=\mathbf{k}^{\prime}-\mathbf{k} + \mathbf{G}$ with $\mathbf{G}$ being the reciprocal lattice vector to take care of the umklapp processes in the crystal.   The momentum dependence of the free matrix element $F_{\rm{DM}}(q)=\alpha_{EM}^2 m_e^2 c^4/(\omega_{\mathbf{k}\mathbf{k}^{\prime}}^2 - \hbar^2 \vert \mathbf{q} \vert^2 c^2)$, where $c$ is the speed of light and $\omega_{\mathbf{k}\mathbf{k}^{\prime}} = E_{+,\mathbf{k}^{\prime}} - E_{-,\mathbf{k}}$. The transition form factor $f_{\mathbf{k}, \mathbf{k}^{\prime}}^{-+} = u_{\mathbf{k}^{\prime},+}^{*} u_{\mathbf{k},-} + v_{\mathbf{k}^{\prime},+}^{*} v_{\mathbf{k},-}$. Here the eigenvalues and corresponding eigenfunctions for the Weyl-Hubbard model are given in equations~(\ref{eq:eigenval-S})-(\ref{eq:eigenfuncm-S}). The halo integral $\eta(v_{\rm{min}}) = \int d^3 v [g_{\chi}(v)/v] \theta(v-v_{\rm{min}})$, where $g_{\chi}(v)$ is the DM velocity distribution and $v_{\rm{min}} =\frac{\omega_{\mathbf{k}\mathbf{k}^{\prime}}}{ \hbar \vert \mathbf{q} \vert} + \frac{ \hbar \vert \mathbf{q} \vert}{2m_{\chi}}$. In the standard halo model~\cite{AKDrukier:1986}, the velocity distribution function for DM is a Gaussian in the Earth's frame, that is, 
\begin{equation}
g_{\chi}(v) = \frac{1}{N_{R,\text{esc}}}  e^{- \frac{\vert \mathbf{v}+\mathbf{v}_{E}\vert^2}{v_0^2}} \theta(v_{\text{esc}} - \vert  \mathbf{v}+ \mathbf{v}_E\vert)\;,
\end{equation}
where the normalization constant $N_{R,\text{esc}}=v_0^3 \pi [\sqrt{\pi} \text{erf}(\frac{v_{\text{esc}}}{v_0}) -2 (v_{\text{esc}}/v_0) \exp[-(v_{\text{esc}}/v_0)^2]$.
This leads to an analytical expression for the halo integral~\cite{REssig:2016}
\begin{subequations}
\begin{equation}
\eta(v_{\text{min}}) = \frac{v_0^2\pi}{2v_E N_{R,\text{esc}}} \biggl{\{} -4 e^{-v_{\text{esc}}^2/v_0^2}v_E + \sqrt{\pi} v_0 \biggl{[} \text{erf}\biggl{(} \frac{v_{\text{min}} + v_E}{v_0}\biggr{)}  - \text{erf}  \biggl{(} \frac{v_{\text{min}} - v_E}{v_0}\biggr{)} \biggr{]} \biggr{\}} \;,
\end{equation}
for $v_{\text{min}} < v_{\text{esc}} - v_E$, while 
\begin{equation}
\eta(v_{\text{min}}) =  
\frac{v_0^2\pi}{2v_E N_{R,\text{esc}}} \biggl{\{} -2 e^{-v_{\text{esc}}^2/v_0^2}(v_{\text{esc}}-v_{\text{min}} +v_E) 
+ \sqrt{\pi} v_0 \biggl{[} \text{erf}\biggl{(} \frac{v_{\text{esc}}}{v_0}\biggr{)}  - \text{erf}  \biggl{(} \frac{v_{\text{min}} - v_E}{v_0}\biggr{)} \biggr{]} \biggr{\}} \;,
\end{equation}
for $ v_{\text{esc}} - v_E < v_{\text{min}} < v_{\text{esc}}  + v_E$.
\end{subequations}
 
The absorption rate of the DM is given by~\cite{YHochberg:2018} 
\begin{equation}
R_{abs} = \frac{\rho_{\chi}c^2 \varepsilon^2}{3\rho_{T}\hbar} \sum_{i} \frac{ \rm{Im}[\epsilon_{i}(m_A)]}{\vert \epsilon_{i}(m_A) \vert^2} \;.
\end{equation}
where  $\varepsilon$ is the kinetic mixing parameter,  $m_A$ is the mass of kinetically mixed dark photon, and the summation is over the three eigenvalues of the dielectric function tensor.

For the numerical calculation of the detection rates, we use the velocity parameter values~\cite{REssig:2016}, $v_0 = 230 $ km/s, $v_{E}=240$ km/s, $v_{\text{esc}}=600$ km/s, respectively, while  the typical dark matter and target mass density~\cite{YHochberg:2018}, $\rho_{\chi}=0.4$ GeV/cm$^3$ and $\rho_{T}=10$ g/cm$^3$, respectively. A 1 kg-year exposure is taken.

\section{Dynamical conductivity and binning method}
The central quantity entering into both the scattering and absorption rates of DM is the momentum transfer dependent dynamical dielectric function 
$\overleftrightarrow{\bm{\epsilon}}(\mathbf{q},\omega)$. This quantity is related to the dynamical conductivity tensor~\cite{MDressel:2002} as $\epsilon_{\alpha\beta}(\mathbf{q},\omega)=
\delta_{\alpha\beta}+\frac{4\pi i}{\omega}\sigma_{\alpha\beta}(\mathbf{q},\omega)$.
For the given model Hamiltonian Eq.~(1) (in the main text) within the Gutzwiller approximation, the conductivity can be calculated directly through the current-current correlations~\cite{JXZhu:2016}  
\begin{equation}
\label{eq:sigma1-S}
\sigma_{\alpha\beta}(\mathbf{q},\omega) =  \frac{1}{i\omega} \biggl{[} -\Pi_{\alpha \beta}(\mathbf{q},\omega) + \frac{e^2 a^2}{\hbar^2} K_{\alpha} \delta_{\alpha\beta}\biggr{]}\;,
\end{equation}
where the current-current correlation function 
\begin{equation}
\Pi_{\alpha\beta}(\mathbf{q},\omega) = \frac{\hbar e^2}{N_L a^3} \sum_{\mathbf{k}^{\prime},ll^{\prime},ss^{\prime}s^{\prime\prime}s^{\prime\prime\prime}} v_{\mathbf{k}^{\prime},ss^{\prime}}^{\alpha}(\mathbf{q}) v_{\mathbf{k}^{\prime}+\mathbf{q},s^{\prime\prime}s^{\prime\prime\prime}}^{\beta}(-\mathbf{q}) \phi_{l\mathbf{k}^{\prime}s^{\prime\prime\prime}}  
\phi_{l\mathbf{k}^{\prime}s}^{*}   \phi_{l^{\prime}\mathbf{k}^{\prime}+\mathbf{q}s^{\prime}} 
 \phi_{l^{\prime}\mathbf{k}^{\prime}+\mathbf{q}s^{\prime\prime}}^{*} \frac{f(E_{l\mathbf{k}^{\prime}}) - f(E_{l^{\prime}\mathbf{k}^{\prime}+\mathbf{q}})}{\hbar \omega+ E_{l\mathbf{k}^{\prime}} -E_{l^{\prime}\mathbf{k}^{\prime}+\mathbf{q}} + i\eta } \;,
\end{equation}
with $\eta$ is the infinitesimal, and the kinetic energy flow
\begin{equation}
K_{\alpha} = \frac{1}{N_L a^3} \sum_{\mathbf{k}^{\prime},l,ss^{\prime}} \kappa_{\mathbf{k}^{\prime}ss^{\prime}}^{\alpha} \phi_{l\mathbf{k}^{\prime}s}^{*} \phi_{l\mathbf{k}^{\prime}s^{\prime}} f(E_{l\mathbf{k}^{\prime}})\;.
\end{equation}
Here $\phi_{l\mathbf{k}s}$ is the components of eigenfunctions given in Eqs.~(\ref{eq:eigenfuncp-S}) and (\ref{eq:eigenfuncm-S}).
The  velocity factors are given by
\begin{subequations}
\begin{eqnarray}
v_{\mathbf{k}ss^{\prime}}^{x}(\mathbf{q})  &=&  \frac{i\alpha t a}{\hbar} [\sigma_{x,s^{\prime}s} e^{-ik_x a} - \sigma_{x,ss^{\prime}}e^{i(k_x + q_x)a}]\;, \\
v_{\mathbf{k}ss^{\prime}}^{y}(\mathbf{q}) &=& \frac{i\alpha t a}{\hbar} [(\sigma_{x,s^{\prime}s} + it^{\prime}\sigma_{y,s^{\prime}s}/t) e^{-ik_y a}  \nonumber \\
&& - (\sigma_{x,ss^{\prime}} + it^{\prime} \sigma_{y,ss^{\prime}}/t) e^{i(k_y + q_y)a} ]\;,\\
v_{\mathbf{k}ss^{\prime}}^{z}(\mathbf{q}) &=& \frac{i\alpha t a}{\hbar} [(\sigma_{x,s^{\prime}s} \textcolor{red}{-} it^{\prime}\sigma_{z,s^{\prime}s}/t) e^{-ik_z a}  \nonumber \\
&& - (\sigma_{x,ss^{\prime}} + it^{\prime} \sigma_{z,ss^{\prime}}/t) e^{i(k_z + q_z)a} ]\;,
\end{eqnarray}
\end{subequations}
and the directional kinetic energy factors are given by
\begin{subequations}
\begin{eqnarray}
\kappa_{\mathbf{k}ss^{\prime}}^{x} &=& -2\alpha t  \sigma_{x,ss^{\prime}} \cos k_x  a\;, \\
\kappa_{\mathbf{k}ss^{\prime}}^{y} &=& -2 \alpha [t  \sigma_{x,ss^{\prime}}  \cos k_y  a -  t^{\prime}   \sigma_{y,ss^{\prime}} \sin k_y a ]\;,\\
\kappa_{\mathbf{k}ss^{\prime}}^{z} &=& -2\alpha [t \sigma_{x,ss^{\prime}} \cos k_z  a \textcolor{red}{-}  t^{\prime}   \sigma_{z,ss^{\prime}} \sin k_z a]  \;.
\end{eqnarray}
\end{subequations}
Equation~(\ref{eq:sigma1-S}) can be recast into a compact form as 
\begin{equation}
\sigma_{\alpha\beta}(\mathbf{q},\omega) = 
\frac{\hbar e^2}{iN_L a^3} \sum_{\mathbf{k}^{\prime},ll^{\prime},ss^{\prime}s^{\prime\prime}s^{\prime\prime\prime}}
\mathcal{C}_{\alpha\beta}^{ll^{\prime},ss^{\prime}s^{\prime\prime}s^{\prime\prime\prime}}
(\mathbf{q},\mathbf{k}^{\prime})
  \frac{f(E_{l\mathbf{k}^{\prime}}) - f(E_{l^{\prime}\mathbf{k}^{\prime}+\mathbf{q}})}{\hbar \omega+ E_{l\mathbf{k}^{\prime}} -E_{l^{\prime}\mathbf{k}^{\prime}+\mathbf{q}} + i\eta } \;,
\end{equation}
where 
\begin{equation}
\mathcal{C}_{\alpha\beta}^{ll^{\prime},ss^{\prime}s^{\prime\prime}s^{\prime\prime\prime}}
(\mathbf{q},\mathbf{k}^{\prime})=\frac{
v_{\mathbf{k}^{\prime},ss^{\prime}}^{\alpha}(\mathbf{q}) v_{\mathbf{k}^{\prime}+\mathbf{q},s^{\prime\prime}s^{\prime\prime\prime}}^{\beta}(-\mathbf{q}) \phi_{l\mathbf{k}^{\prime}s^{\prime\prime\prime}}  
\phi_{l\mathbf{k}^{\prime}s}^{*}   \phi_{l^{\prime}\mathbf{k}^{\prime}+\mathbf{q}s^{\prime}} 
 \phi_{l^{\prime}\mathbf{k}^{\prime}+\mathbf{q}s^{\prime\prime}}^{*}}{E_{l\mathbf{k}^{\prime}} - E_{l^{\prime}\mathbf{k}^{\prime}+\mathbf{q}}} \;.
\end{equation}

By inserting an integral over the Dirac delta function, we find
\begin{equation}
\sigma_{\alpha\beta}(\mathbf{q},\omega) = 
 \int_{-\infty}^{\infty}dx ~
 \frac{\tilde{\Pi}^{\alpha\beta}(\mathbf{q},x)}{\hbar \omega-x+i\eta},
 \end{equation}
 where 
 \begin{equation}
\tilde{\Pi}^{\alpha\beta}(\mathbf{q},x)= 
\frac{\hbar e^2}{iN_L a^3} \sum_{\mathbf{k}^{\prime},ll^{\prime},ss^{\prime}s^{\prime\prime}s^{\prime\prime\prime}}
\mathcal{C}_{\alpha\beta}^{ll^{\prime},ss^{\prime}s^{\prime\prime}s^{\prime\prime\prime}}
(\mathbf{q},\mathbf{k}^{\prime})
 \biggl{[} f(E_{l\mathbf{k}^{\prime}}) - f(E_{l^{\prime}\mathbf{k}^{\prime}+\mathbf{q}})\biggr{]}
  \delta\left(x-\left[E_{l^\prime{\mathbf{k}^\prime}+\mathbf{q}}-E_{l\mathbf{k}^{\prime}}\right]\right) \;.  
 \end{equation}
 In our numerical calculations, we use the Dirac identity 
\begin{equation}
\lim_{\eta \rightarrow 0} \frac{1}{\hbar \omega-x+i\eta}=\mathcal{P}\frac{1}{\hbar\omega-x} -i\pi \delta(\hbar\omega-x)\;,
\end{equation}
which leads to 
\begin{equation}
\sigma_{\alpha\beta}(\mathbf{q},\omega) = 
 \mathcal{P}\int_{-\infty}^{\infty}dx ~
 \frac{\tilde{\Pi}^{\alpha\beta}(\mathbf{q},x)}{\hbar \omega-x} -i\pi \tilde{\Pi}^{\alpha\beta}(\mathbf{q},\omega)\;.
\end{equation}
In this way, the calculation of $\sigma_{\alpha\beta}$ is split into two parts: (i) a binning process over the excitation energies $E_{l^\prime{\mathbf{k}^{\prime}+\mathbf{q}}}-E_{l{\mathbf{k}^{\prime}}}$ for a given $\mathbf{q}$, and (ii) a principal value integral aided with the modified Gaussian rule
 to recover $\sigma_{\alpha\beta}$. This decoupling reduces the computational complexity from $N_{\mathbf{q}} \times N_{\mathbf{k}} \times N_{\omega}$ to $N_\mathbf{q} (N_\omega N_X + N_\mathbf{k})$, and yields a significant computational advantage when $N_X <N_{\mathbf{k}}$, which is usually the case. This scheme enables us to undertake calculations on dense $\mathbf{q}(\mathbf{k})$ and $\omega$ meshes.
 We note that our developed formalism contains the contribution from the entire momentum space in the Brillouin zone, which goes beyond the earlier studies~\cite{YHochberg:2018} with the phase space confined to the Dirac cone regions.